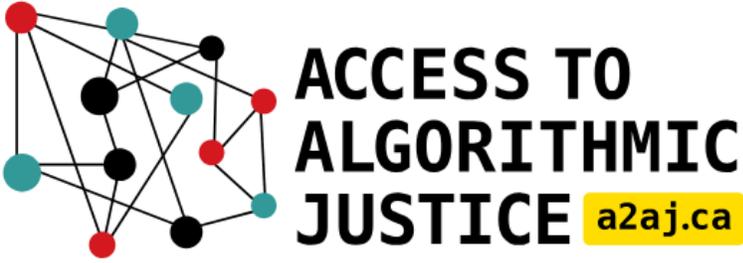

# Access to Algorithmic Justice Working Paper

Introducing the A2AJ's *Canadian Legal Data*: An open-source alternative to CanLII for the era of computational law[⊗]

*Simon Wallace[*] & Sean Rehaag[**]*

September 15, 2025**Abstract** The Access to Algorithmic Justice project (A2AJ) is an open-source alternative to the Canadian Legal Information Institute (CanLII). At a moment when technology promises to enable new ways of working with law, CanLII is becoming an impediment to the free access of law and access to justice movements because it restricts bulk and programmatic access to Canadian legal data. This means that Canada is staring down a digital divide: well-resourced actors have the best new technological tools and, because CanLII has disclaimed leadership, the public only gets second-rate tools. This article puts CanLII in its larger historical context and shows how long and deep efforts to democratize access to Canadian legal data are, and how often they are thwarted by private industry. We introduce the A2AJ's Canadian Legal Data project, which provides open access to over 116,000 court decisions and 5,000 statutes through multiple channels including APIs, machine learning datasets, and AI integration protocols. Through concrete examples, we demonstrate how open legal data enables courts to conduct evidence-based assessments and allows developers to create tools for practitioners serving low-income communities.---

[⊗] This article draws on research supported by the Social Sciences and Humanities Research Council and the Law Foundation of Ontario. Some sections of the article, as well as some parts of the underlying code used for the project, were drafted with the assistance of generative AI.[*] Co-Director (Access to Algorithmic Justice) & Assistant Professor (Lincoln Alexander School of Law, Toronto Metropolitan University)
[**] Co-Director (Access to Algorithmic Justice) & Associate Professor (Osgoode Hall Law School, York University)1

# Table of Contents

# I. INTRODUCTION

The digitalization of legal texts and advances in new computational research techniques, augers an era of computational law. Pattern detection, network analyses, generative artificial intelligence: each of these promise to change how we gather, organize, and understand legal problems – transforming our relationships with law.

But as Canada enters this era, a troubling informational asymmetry is emerging. Just as in previous decades, when large swaths of caselaw were only accessible to lawyers who could afford to purchase access to privately owned collections of law, the benefits of new legal technologies built on artificial intelligence today accrue disproportionately to powerful institutional actors who collect, hold, and protect Canadian legal data. These advantages also flow mostly to those who can afford fees to access that proprietary data.

In the 1990s, law societies across Canada recognized that asymmetric access to digital collections of law caused significant unfairness: lawyers representing clients with deep pockets were able to leverage proprietary electronic legal databases while lawyers for lower income clients were stuck looking up the law in dusty old books, and sometimes the cases cited by the other side weren't even in those books. One important solution to this unfairness was that the Federation of Law Societies established CanLII in 2001, with a mandate to provide online open access to Canadian legal documents, including case law and legislation. As a result, CanLII now holds one of the largest legal datasets in Canada, with over 3 million legal documents – and for many courts and tribunals, CanLII is the only place where decisions are available online free of charge.

Despite its mandate, CanLII is now attempting to prevent anyone from using its large legal dataset to build new legal technologies, relying on both restrictive terms of use that prohibit bulk or programmatic access, as well as claims that reproducing its dataset without permission violates CanLII's copyright over their dataset.

CanLII's refusal to share its legal data in bulk harms the legal sector and the broader public by limiting the pace and scope of legal innovation. It also disproportionately impacts low-income communities and the lawyers that serve them. Due to limited resources, these communities and lawyers cannot purchase expensive legal tech. And developers who would like to build inexpensive – or even free – services for these communities and lawyers cannot afford to purchase licenses to the proprietary legal data needed to build those services without passing along those expenses to people who cannot afford to pay them. Meanwhile, Thomson Reuters, LexisNexis, Harvey, and BlueJ Legal each are building sophisticated products available, of course, for a hefty fee.

This is how, as Canada enters the era of computational law, where access to the best computational legal tools will be a key vector in distributions of access to quality legal services, CanLII has become barrier to access to justice.

In this article, we introduce an open-source alternative to CanLII: the Access to Algorithmic Justice's (A2AJ) Canadian Legal Data. This project provides multiple open-access and open-source channels to access Canadian court cases, legislation and regulation programmatically and in bulk.

The article begins by outlining the history of efforts to increase the public's access to information. Sometimes these efforts were led by private industry and then petered. Sometimes they were government initiatives that failed. Sometimes they were non-profit ventures that were captured by commercial imperatives. The bottom line is clear: increasing access to legal data is not an idea whose time has come, it is an idea whose idea came decades ago. Action is overdue. Next, we describe the A2AJ and our Canadian Legal Data project, explaining what data we have collected, detailing how to access the data programmatically, and demonstrating how the data can be used to advance access to justice. Finally, we conclude with some ideas for the future of projects like this one. This includes the hopeful possibility that CanLII might choose to fulfill its mandate by making legal data freely available for computational use, rendering this project unnecessary and allowing organizations like A2AJ to focus on using legal data to advance access to justice rather than collecting it.

# II. HOW CANLII BECAME A BARRIER TO ACCESS TO LEGAL INFORMATION IN THE COMPUTATIONAL AGE

## Canada's first steps toward public legal information were promising but incomplete

The first law reporters were exactly that: reporters who attended Court and then published their own accounts of the proceedings. Sometimes these reports included quotations from judges. Sometimes they were the reporter's own notes about the proceeding. Sometimes they were full copies of judgements. Legal reporting was haphazard and inconsistent, dependent on the skills, motivations, and interests of individual reporters and editors. Sometimes in nineteenth century Canada a reporter's work was subsidized by the local law society, but sometimes they were entirely private initiatives.[1]

This made it innovative when, in 1875, Parliament required the new Supreme Court of Canada to report its own decisions. This made the Supreme Court Reports (the SCRs) the first official, the first public, and the first institutional reporter in either the Dominion or the United Kingdom.[2]

But although the SCRs were novel, they were not very good. Lawyers complained that the reports were riddled with errors, typographical problems, delivered too late, and got basic of Canadian law concepts wrong.[3] Worse, they were incomplete. In the early days of the Court, judges often sent the Court registrar notes to explain their votes instead of what we would recognize today as a formal judgement, and sometimes the quality of these judicial notes was uneven.[4]

But a judgement explained poorly is better than no explanation at all and the early SCRs struggled to understand why judges voted one way or another. Sometimes the SCRs were delayed because judges took months to forward their explanatory materials to the reporter.

---

[1] Vivienne Denton, "Canadian Law Publishers: A look at the development of the legal publishing industry in Canada" in Martha Foote, ed, *Law Reporting and Legal Publishing in Canada: A History* (Canadian Association of Law Libraries, 1997) at 17. [Canadian law publishers]
[2] James Snell and Frederick Vaughn, *The Supreme Court of Canada: History of an Institution* (Toronto: The Osgoode Society, 1985) at 35.
[3] Ibid at 36.
[4] Ibid at 37.

Other times, judges just did not send their judgements at all. In the report for *Milloy v. Kerr*, for example, the official text says that Justice Strong prepared "a written judgment in favor of affirming the judgment of the Court of Appeal."[5] But where the text is supposed to be, there is only a footnote that explains "[t]he learned judge, having mislaid his judgment, directed the reporter to report the case without it."[6]

Was this new public publishing effort a threat to private publishing ventures? Clearly no. Throughout the nineteenth and twentieth century, new legal publishers established themselves and succeeded. Many major publishers today have roots in early Canadian history. For example, Canada Law Book grew out of the *Upper Canada Law Journal* (1855) and would go on to publish the *Canada Law Journal*, the *Canadian Criminal Cases*, and the *Dominion Law Reports*.[7] Robert Carswell owned a series of Toronto bookstores in the late ninetieth century, before shifting to focus on publishing law books.[8] Americans, too, took an interest. Commerce Clearing House (CCH) was a Chicago-based publisher that started in 1913 and, by the 1950s, was publishing an extensive collection of Canadian law materials.[9] To be sure, much of the legal publishing industry was centralized in Toronto but as the example of Maritime Law Book (established in the 1960s) shows, there were markets for private publishers outside of central Canada.[10]

## Academic innovation in legal technology was quickly captured by commercial interests

As the private publishing industry grew, technological disruption loomed. In the 1960s, out of the limelight, research and development in computational databases and information retrieval methods accelerated.[11] Part of this was driven by (ironically or appropriately) legal factors. The American Department of Justice initiated a series of antitrust investigations into IBM in the late 1960s, which required the company to develop systems to organize and analyze huge masses of unstructured text data.[12] Quickly, IBM realized that this search and database work could be a new product line for large business.

Uptake was not immediate, and a lot of the early experimental work happened in university labs. Professor Hugh Langford at the Queen's University law school was one of the international leaders in the field. Originally interested in finding ways to catalogue and organize treaties, in the 1960s he partnered with IBM – which gave free access to some its database tools and assigned staff to the university[13] – and the Canadian Department of Justice to establish the Queen's University Institute for Computing and Law (QUIC/LAW). The purpose of the lab was to explore how novel computational technologies could allow people to access legal information more quickly.[14]

---

[5] Ibid and *Milloy v. Kerr*, 1880 CanLII 5 (SCC), 8 SCR 474.
[6] Ibid.
[7] Canadian law publishers, *supra* note 1, at 22.
[8] Ibid at 24-26.
[9] Ibid at 29-31.
[10] Ibid at 36.
[11] Charles Bourne and Trudi Bellardo Hahn, *A History of Online Information Services, 1963–1976* (Cambridge: MIT Press, 2003) at 1.
[12] Ibid at 127 and ff.

The young Langford would today be an open access to law activist. He thought that the "whole point of computerized information systems is to enable more people to have access to more information quickly."[15] His work in the university mattered, he thought, because he was in a race against the private sector that would treat search technology as a "valuable commodity and to restrict rather than expand its use."[16] And if these were new views, they were not isolated ones: similar projects were cropping up all over North America.[17]

At first, the work of QUIC/LAW was extraordinarily labour intensive. The initiative hired over 60 people to type up laws, cases, and books so that they could be entered into digital (then tape) databases. Once the digital databases were large enough, researchers used their new search technology to enable plain language searches of caselaw. Early pilot projects involved the installation of large terminals in Department of Justice and some private law offices. The other major QUIC/LAW innovation concerned text editing. As Parliament and the legislatures amended laws, QUIC/LAW developed approaches that allowed for quick updating and consolidation of law. Soon, QUIC/LAW had a contract with the Speaker of the House of Commons to help produce up-to-date versions of orders and Hansard.[18]

Funding for QUIC/LAW dried up. The partnership with IBM ended in 1969[19] and, by 1973, the Department of Justice pulled its support.[20] Langford, however, was not deterred he spun off his research into a new private venture: QL systems. Langford and his partner, Professor Richard von Briessen, agreed to purchase QUIC/LAW from Queen's University for $130,000, in payments spread out over seven and a half years.[21] This was, we can now see clearly, money well spent but it was a development rich with irony. As Mark Walters ruefully observes, the point of this new private project was to "pursue what they feared someone else would pursue – the commercial computerization of the law."[22]

---

[13] Ibid at 125.
[14] Mark Walters, "Let Right Be Done: A History of the Faculty of Law at Queen's University" (2007) 31 Queen's LJ 213 at 345. [Let Right be Done]
[15] Ibid, cited at 345.
[16] Ibid.
[17] The First National Conference on Automated Law in 1972 just under 300 participants: Ronald May, "First National Conference on Automated Legal Research: a Report for Jurimetrics" (1972) 13 Jurimetrics 3.
[18] "Putting the law into the computer" Globe and Mail (14 May 1974) 8.
[19] Charles Bourne and Trudi Bellardo Hahn, *A History of Online Information Services, 1963–1976* (Cambridge: MIT Press, 2003) at 314.
[20] Let Right be Done, supra note 14, at 345.
[21] Hugh Winsor, "$2.5 million system for computers sold to researchers: Ottawa pays but finders keepers" Globe and Mail (14 May 1974) A8.
[22] Ibid.

The reasons why the Department of Justice withdrew its funding are not clear. Perhaps it is because the National Research Council developed its own search tool, CAN-OLE, and the QUIC/LAW project was redundant. Or, perhaps, government was worried about the larger social consequences of search technology. In 1974, the Department of Communications published a report prepared by Professor Phillip Slayton of McGill University about the potential impacts of electronic retrieval of legal information. Electronic access to legal information could revolutionize, he thought, the practice of law, but not necessarily for the better. He worried that the very foundation of the common law was at risk. Instant access to the full scope of potential legal information could eliminate judicial creativity because judges would no longer be called upon to make law, for every possibility would, somewhere in the great sea of data, have a precedent. Worse, electronic search might "accentuate existing social inequalities by providing superior legal information" to those who could pay.[23] These risks led Layton to call for a moratorium on public funds for legal search technology.

But, as they say, the cat was out of the bag and Langford did not need public money to succeed. By the 1980s QL Systems was a computational winner. Langford licensed the core search technologies behind QL to West Publishing Company by the mid 1970s, which in turn used it to develop the Westlaw database.[24] Canada Law Book also took a significant stake in the company that decade.[25] In 1999 QL was reincorporated as Quicklaw and then, in 2004, it merged with LexisNexis Butterworths Canada.[26] If today computerized search of law is ubiquitous, it is no small part because of the work of a university professor in Kingston in the 1960s.

To be sure, QuickLaw did not have and does not have a complete monopoly in the search space. In Québec, DATUM – a similar project to QUIC/LAW – started around the same time, but the lab remained in public hands. DATUM was also an experiment to see how computers could assist with the retrieval of legal information and help lawyers do research, but it was quickly folded in a state-based non-profit. In the 1970s, the National Assembly of Québec established the Société québécoise d'information juridique (SOQUIJ). This non-profit organization was given official responsibility for collecting and disseminating Québec law. In a model that would anticipate CanLII, each lawyer in the province contributed to SQUIJ with the province covering any deficit.[27]

---

[23] Phillip Slayton, *Electronic Legal* Retrieval (Ottawa: Department of Communications, 1974) at 25.
[24] Ibid at 346 and James Sprowl, "The Westlaw System-A Different Approach to Computer-Assisted Legal Research" (1975) 16 Jurimetrics 142.
[25] Ejan Mackaay, "User preferences, experiments and the question of the initiative in automated law retrieval in Canada" (1977) 8:1 Revue de droit de l'Université de Sherbrooke 97 at 99. [Automated law retrieval in Canada]
[26] "LexisNexis Butterworths Canada Acquires Quicklaw" *Lexpert* online: https://www.lexpert.ca/big-deals/lexisnexis-butterworths-canada-acquires-quicklaw/343383#:~:text=On%20July%2015%2C%202002%2C%20Butterworths,and%20Peter%20Cooke%20(trademarks).
[27] Automated law retrieval in Canada, supra note 25 at 103.

A similar initiative to develop a public legal information repository for the common law provinces was dead on arrival. In 1975 the Canadian Law Information Council (CLIC), an organization that received most of its funding from the Department of Justice, proposed to develop a case retrieval service in Toronto, based on the Quebec model. But the project floundered because the English language publishers would not share their databases, a fact which could be "interpreted only as a desire to reserve for themselves the right to exploit a computer retrieval service, if it was at all commercially promising."[28]

CLIC never entirely abandoned its ambition to make a public and free search tool, but its efforts proved fruitless. At a 1986 meeting, for example, CLIC resolved to support the "concept" of an "integrated distribution system of decisions (a clearinghouse concept)" and specifically encouraged courts and tribunals to "provide free and equal access to their decisions in all available forms to all publishers, information suppliers, and the public" and "encourage[d] publishers… who gain electronic access to decisions… to provide electronic distribution… to other publishers and suppliers."[29] Finally, in 1992, the federal government decided the issue by withdrawing its funding.[30] Likely, CLIC was a minor casualty of the larger Chretien era austerity agenda, but surely there must also be a kernel of truth to former Alberta Court of Appeal Justice Jean Côté's claim that the "Toronto publishers suspected the Council of socialist tendencies, and it got disbanded."[31]

## The Internet era promised democratic access to law but created new gatekeepers

Up until the 1990s, anyone could be forgiven for not caring much about electronic search. Even if QL was technologically impressive, users needed a specialized computer terminal to access its data and search its decisions. Throughout the 1960s, 1970s, and 1980s, electronic search was a decidedly scholarly, enterprise, or big government technology. The proliferation of CD-ROM technology in the 1990s democratized, in one limited sense, access to search as publishers started to distribute electronic texts, books, and databases direct to legal consumers. Then, by the late 1990s, publishers began to sell access to online subscriptions.

These products were not cheap. Beginning in 1997 the *Canadian Law Libraries* journal started to publicly track the proliferation of digital (CD-ROM and internet) research products and report on their costs.[32] Prices varied but were generally significant: in 1999, the eCarswell "Securties.pro" module (for firms with 15+ lawyers) cost $7,396.08 and its "Family.Pro" module cost $1,595 per user, while Maritime Law Book charged non-subscribers $1.50 to access a single case online.[33]

---

[28] Ibid at 115-116.
[29] "Report of the Canadian Law Information Council Meeting (21 November 1986)" (1986) Canadian Association of Law Libraries Newsletter-Bulletin 212 at 212-213.
[30] Tom McMahon, "Improving Access to the Law in Canada With Digital Media" (1999) 16 Government Information in Canada/Information gouvernementale au Canada, online: http://library.usask.ca/gic/16/mcmahon.html.
[31] Jean Côté, "Access to Court Decisions" (15 Sept 2017), online: https://www.legalviews.com/coteopinion10.htm.

In the 1990s the Federal government got into the game. Against the growing private tendency to charge for access to data, the federal government decided to orient to the opportunities of informational exchange on the internet differently. In its 1995 final report, the *Information Highway Advisory Council* recommended that the federal government "as a rule, place federal government information and data in the public domain."[34] The next year, the government enacted the *Reproduction of Federal Law Order*, which allowed "anyone... without charge or request for permission" to "reproduce enactments and consolidations of enactments of the Government of Canada, and decisions and reasons for decisions of federally-constituted courts and administrative tribunals" subject only to a due diligence requirement.[35] By 1995, it was also publishing all Canadia laws (save the *Income Tax Act* and the *Customs Act*) online,[36] and made a CD-ROM version of its laws available for $225.[37]

Federal tribunals and courts, too, got on board. In 1993, a new university-based access-to-legal-information lab was born when Daniel Poulin started LexUM. If QUIC/LAW and DATUM were of the first digital search moment, LexUM was of the early Internet moment. For its first major project, it built a Gopher site (Gopher was an early rival to the World Wide Web) for the Supreme Court of Canada. Over the 1990s, it continued to develop web-based tools to publish law online, often using the revenue from contracts with tribunals to fund its research and scholarly work.[38]

---

[32] Sue Bengin, "Tracking the Cost of Canadian Legal Subscriptions – 1997" (1998) 23:1 Can L Libr 40 at 40.
[33] Sue Bengin, "Tracking the Cost of Canadian Legal Subscriptions – 1999" (1999) 24:1 Can L Libr 36 at 37.
[34] Information Highway Advisory Council, *The Challenge of the Information Highway: Final Report of the Information Highway Advisory Council* (Ottawa, 1995) at 117.
[35] *Reproduction of Federal Law Order*, SI/97-5.
[36] Tom McMahon, "Improving Access to the Law in Canada With Digital Media" (1999) 16 Government Information in Canada/Information gouvernementale au Canada, online: http://library.usask.ca/gic/16/mcmahon.html.
[37] See *Tolmie v. Canada (Attorney General) (T.D.)*, 1997 CanLII 5545 (FC), [1997] 3 FC 893.
[38] Daniel Poulin, "CanLII: How Law Societies and Academia can Make Free Access to Law a Reality" (2004) 1 The Journal of Information, Law and Technology. [Law societies and academia]

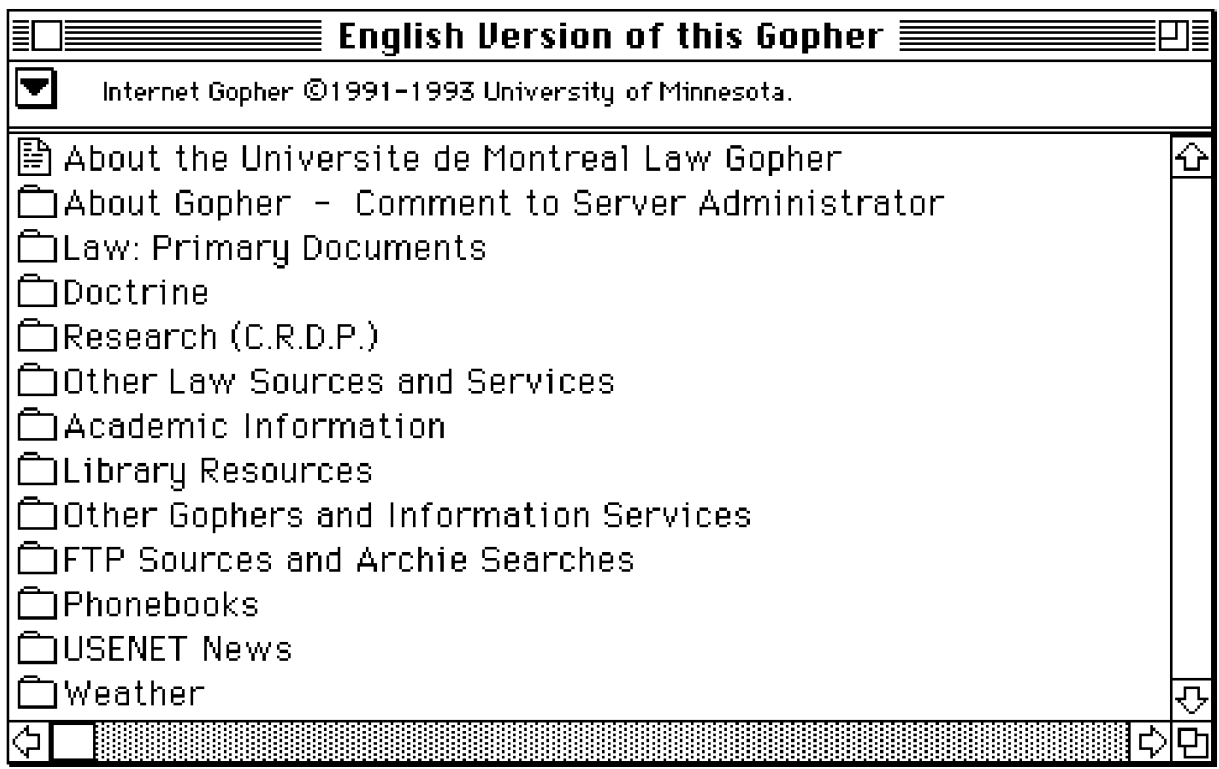

*Figure 1 LexUM's Gopher Site for free access to law, online: https://web.archive.org/web/20061024215006/http://www.lexum.com/doc/static/gopher.1993.09.pdf*

And if the federal space looked encouraging, In Québec, the picture initially looked even more salutary. Unlike QUIC/LAW, DATUM (the similar project based out of the University of Montreal) was never formalized, becoming part of SOQUIJ. Over time, the SOQUIJ expanded its data offerings and, by the 1990s, was publishing roughly 20% of all decisions released by Québec courts.[39]

For private publishers, this presented a problem: they wanted all of the decisions so they could publish their own offerings, but SOQUIJ was preventing them obtaining access to the decisions. Corporate activism was led by the publisher Wilson & Lafleur. It independently negotiated an agreement with the Québec Court of Appeal to receive all of that Court's decisions for an annual bulk fee of $1,750 and a separate agreement for all "Quebec court decisions affecting the interpretation of the Civil Code of Quebec from SOQUIJ for $0.34 per page."[40] But the remainder of unpublished cases were out of private entities reach.

In the mid-1990s, the publisher approached SOQUIJ and asked it to provide copies of all the remaining decisions. SOQUIJ refused and Wilson & Lafleur launched a civil action. In 2000, the publisher was vindicated by the Québec Court of Appeal, which found that Wilson &

---

[39] *Wilson & Lafleur inc. v. Société québécoise d'information juridique*, 2000 CanLII 8006 (QC CA), para 32.
[40] Ibid, para 11.

Lafleur "has (as does everyone) the right to access the full text of all decisions handed down by Quebec courts, a right that has been recognized by the Quebec legislature and enshrined in the Canadian constitution."[41] It ruled that "SOQUIJ must therefore provide Wilson & Lafleur access to all judgments coming out of Quebec courts to which it itself has access."[42]

This seemed like a good outcome. One commentator explained that instantly "Quebec went from a place where free publication of law had made painfully little progress to eventually become the Canadian jurisdiction where case law is most accessible." Shortly after the judgement the "Quebec government adopted a new policy and mandated that the Société québécoise d'information juridique set up a Web site offering a free basic access to all decisions rendered by courts and tribunals in Quebec."[43] But time can atrophy good things. When the A2AJ asked for access to SOQUIJ's database, the publisher refused explaining that bulk access to caselaw is only for CanLII, WestLaw, Quicklaw, and the CAIJ.[44]

In Ontario, the roles reversed: private interests sought to limit access to legal information while semi-public bodies worked to increase access. In 1993, three legal publishers (CCH Canadian Ltd., Thomson Canada Ltd. and Canada Law Book Inc.) sued the Law Society of Upper Canada (LSUC). Through its library, the LSUC provided a "request-based photocopy service" that allowed certain library users to ask the staff to photocopy specific materials and send it to them by mail or by fax. The publishers said that this program infringed their copyright. Eleven years later, the Supreme Court of Canada resolved the case in favour of the LSUC, explaining that it did not breach contract by sharing photocopies of legal texts.[45] This was, to be sure, a victory for advocates of the free dissemination of law, but its analogue conclusions arrived too late to seriously influence the emerging digital world.

The Federal of Law Society of Canada decided to get out in front of the issue. It struck a *National Virtual Law Library Group* to explore the viability of setting up a pan-Canadian organization, similar to SOQUIJ, to collect and disseminate primary legal materials. In March 2000, the group released its core report, arguing that there were a range of options available to the Federation: it could purchase an existing publisher, start a new non-profit, or establish some form of new institute. For the report drafters, these were heady plans. At the end of the report they recalled that the "seekers of the treasure of Sierra Madre had deserts and mountains to overcome in search of mythical treasure. The difference in the case of the Virtual Law Library is that the treasure of the Virtual Law Library is indeed within our reach."[46]

---

[41] Ibid, para 1.
[42] Ibid, para 34.
[43] Email on file with authors.
[44] https://citoyens.soquij.qc.ca/editeurs/
[45] *CCH Canadian Ltd. v. Law Society of Upper Canada*, 2004 SCC 13 (CanLII), [2004] 1 SCR 339.
[46] David Brusegard, *Toward a Business Plan for a Canadian Virtual Law Library* (Montreal: Federal of Law Societies, 2000) at 43.

## LexUM embraces market imperatives and so does CanLII

Ultimately, the Federation decided to create an institute that would be partly financed by a levy paid by each lawyer in Canada. Part of the reason that the Federation was confident that it could get the project off the ground is because of LexUM said it would partner with the Federation. The agreement was simple: "LexUM would build CanLII" provided that the Federation respect LexUM's academic mandate. As Daniel Poulin later explained, in a paper titled "How Law Societies and Academia Can Make Free Access to Law a Reality", "[t]he goals of the academics involved in CanLII were clear: to increase access to law, to develop Canadian know-how in the subject field, to create a place where research on computerized legal documents can be conducted and to offer training for students."[47]

The decision to create CanLII immediately raised LexUM's international stature. CanLII was not the first Legal Information Institute (Cornell Law School's Legal Information Institute is the undisputed first),[48] but CanLII quickly became one of the most successful free legal information institutions in the world. In 2002, LexUM hosted a gathering of legal information institutes from around the world to sign the *Montreal Declaration on Free Access to Law* that, amongst other things, said that "[p]ublic legal information is digital common property and should be accessible to all on a non-profit basis and free of charge."[49]

In Canada, CanLII disrupted the marketplace. By 2001, CanLII hosted over 325,000 files over forty different collections, allowing anyone with an internet connection to access individual files.[50] A few years before, this was barely imaginable. Before CanLII, the established publishers would explain "to all who wanted to listen, and to many who did not, that publishing law was so costly that they could not charge less than a couple of hundred dollars an hour for use of their systems."[51] CanLII's public access model put-paid to that claim. Of course, there remained resistance. As Michael Geist, a board member of CanLII explained to Parliamentarians in 2010 not everyone wanted to participate: "some provinces even see access to things like provincial statutes as a potential revenue opportunity and thus create restrictions for people to be able to access that information."[52]

Regardless, CanLII grew and, in keeping with its academic mandate, LexUM was able to develop new computational tools and new ideas about how law ought to be accessed. In 2008, Poulin proposed a development to WorldLII. The plan here was bold. He proposed that each LII who wanted to participate in the program support a standard application programming interface (a means to share information between hubs ) to allow a user to interact with any LII around the world, creating a federated LII.[53] LexUM also developed innovative and useful technologies. For example, using the CanLII dataset, LexUM developed the Reflex citator,[54] a tool that helped map the connections between different judgements.

---

[47] Law societies and academia, *supra* note 33.
[48] Daniel Poulin, "Fifteen Years of Free Access to Law" (Lexum, 2008) at 11, online: https://lexum.com/en/blog/fifteen-years-of-free-access-to-law. [Fifteen years]
[49] *Declaration on Free Access to Law* (Montreal, 2022).
[50] Law societies and academia, *supra* note 33.
[51] Fiteen years, *supra* note 48.
[52] Canada, House of Commons, Standing Committee on Access to Information, Privacy and Ethics, *Minutes of Proceedings and Evidence*, 40th Parl, 3rd Sess, No 38 (9 Dec 2010) (Michael Geist) at 1600.

But the academic model eventually, and for reasons that are not entirely clear, came under stress. In 2010, LexUM ended its affiliation with the University of Montreal and became a private company, presumably owned primarily by Poulin. All of this was foreshadowed. When Hugh Langford, the founder of Quic/LAW passed away, Poulin recalled that the two of them often "skirmished." One day, Poulin later recalled, they were having coffee and "Langford told me in his mischievous manner that LexUM will end up as a business. I disagreed without much nuance. Later on, he took the check and we split."[55]

By now, Poulin was arguing that the private industry and profit motives could be harnessed to increase access to legal information. The "business of the for-profit company," he once explained, "is to help make the law accessible, not only as a service provider to CanLII, but also through all the other products it sells."[56]

Lexum pursued this vision through by pursuing two revenue streams. First, it tried to monetize its privileged access to CanLII. For example, in 2015 Lexum released a tool called LexBox that helped users on CanLII keep track of their research. At the time of release, Poulin explained that Lexum hoped "adopt some sort of freemium model, where what is there is for free and something more will call for subscriptions."[57] Little progress was made on this front and, save for a short-lived partnership with Clio, LexBox appears not have succeeded.

Second, LexUM leveraged its work with CanLII not only to get paid for the services it provided, but to fund its product development. The financial arrangements between LexUM and CanLII are opaque, but between 2013 and 2015 CanLII's overall budget rose from $2.8 to $3.1 million, of which, some significant portion of the funds must have been remitted to LexUM.[58] By 2013, Lexum explained that it focused its "its energy on developing standardized products that can be shared among its clientele" and that its "products uniformly address the common needs of legal information providers, but also offer a large selection of optional added-value features giving a unique flavour to every release."[59] Or, to put it simply, it took its work paid for by CanLII and tried to make it attractive to the wider market.

---

[53] Fifteen years, *supra* note 48, at 18-19.
[54] Daniel Poulin, Éric Paré and Ivan Mokanov, "Reflex – Bridging Open Access with a Legacy Legal Information System" (delivered at *Law via the Internet 2005*). See also Janine Miller, "The Canadian Legal Information Institute - a Model for Success" (2006) 8:4 Legal Information Management 280 at 282.
[55] Comment on Simon Chester, "Hugh Lawford 1933-2009" (18 August 2009), online: https://perma.cc/EQ9F-765T
[56] Daniel Poulin, "25 for 25: So it's been twenty-five years. The LII, its descendants, and their future" online: https://blog.law.cornell.edu/voxpop/2017/02/10/25-for-25-so-its-been-25-years-the-lii-its-descendants-and-their-future/
[57] Comment on Nate Russel, "Of Lexbox and the Promise of Convenience for CanLII Users" (16 July 2015), online: https://www.slaw.ca/2015/07/16/of-lexbox-and-the-promise-of-convenience-for-canlii-users/.
[58] *Canadian Legal Information Institute v. The Queen*, 2020 TCC 56 (CanLII), para 11.

Did the incentive to generate revenue divert CanLII and Lexum from their larger access to justice goals? It appears so. Developers and researchers interested in working with CanLII data at scale cannot because of CanLII's terms of service, which prevent individuals from systematically access CanLII's databases.[60]

To be fair, some efforts were made to offer bulk access to a portion of CanLII's data. In 2013, CanLII's President Colin Lachance announced a new CanLII project: a platform allowing developers to computationally access CanLII's data, albeit initially restricted to metadata and not to the full text of decisions. Lachance indicated that these limitations were due to budgetary constraints ("Developing the tools required a commercial agreement between CanLII and Lexum Inc., CanLII's technology supplier, and so our ability to expand the range of what is available is naturally influenced by our ability to pay for the necessary development") and because of certain limitations imposed on CanLII by the Courts when they agreed to share their decisions.[61] Regardless, Lachance was optimistic that the developer platform would expand. "We will get there," he wrote.[62]

But we did not and, by 2015, Lachance was no longer the President of CanLII. A few months later, Lachance wrote a blog post – perhaps hinting at tensions between LexUM and CanLII – arguing that CanLII needed to "support the growing ecosystem of legal information innovators by systematizing and standardizing access terms."[63] Poulin disagreed, replying that CanLII and Lexum "do not have an obligation of giving away their content. They could do it, but they don't have the duty to do it."[64] If someone wanted to get in the free access to data game, he suggested, they needed to establish their own agreements with courts and other legal bodies.

This debate was set against the first major misuse of CanLII's data. CanLII always explicitly forbade anyone from systematically downloading the data from its databases. This policy was breached when in 2014 a Romanian-based website downloaded judgements from CanLII, posted them online, and made them available to search engines for indexing. This was a thinly veiled extortion racket because the website offered to remove judgements from its site when individuals, embarrassed that their most intimate affairs were just a Google search away from discovery, paid a fee. Eventually, the Information Commissioner and the Federal Court of Canada intervened, but it appears that, regardless of what Canadian judicial institutions had to say about a project in Romania, the scheme fizzled.[65]

---

[59] Pierre-Paul Lemyre, *Blending In: Lexum's Approach to Cloud-Based Services* (Lexum, 2013) at 7.
[60] "Terms of Use", online: https://www.canlii.org/info/terms.html.
[61] Colin Lachance, "Unbundling legal information" (21 March 2013), online: https://www.slaw.ca/2013/03/21/unbundling-legal-information/
[62] Ibid.
[63] Colin Lachance, "What Does It Really Mean to "Free the Law"? Part 2" (9 November 2015), online: https://www.slaw.ca/2015/11/09/what-does-it-really-mean-to-free-the-law-part-2/.
[64] Comment on Ibid.

By now, CanLII and LexUM were each established players and LexUM's owners decided it was time to cashout and, in 2018, CanLII purchased LexUM.[66] The law societies provided the financing for a deal that appears to have been worth approximately $4 million.[67] At the same time, LexUM's products appear to have found new success on the market as American tribunals and organizations began to use them. Public disclosure databases show, for example, that Arkansas paid LexUM $62,347.72 (USD) in 2024 and $64,218.15 in 2025[68] for its Supreme Court decision search tool[69] and, between 2019 and 2024, the state of New Mexico paid Lexum over $2.9 million (USD) for its NMOneSource portal.[70] These are just two examples of Lexum's private work and, given that we know that LexUM provides support for many Canadian courts and tribunals as well, we can infer that the mature LexUM product generates significant revenue for CanLII.

LexUM also explored how it could use its access to CanLII's data to develop new artificial intelligence approaches to legal research. In Fall 2018, for example, Lexum issued a call for industry partners to join its new Lexum Lab. Lexum explained that it used "machine learning and deep learning techniques on the millions of documents available on the Canadian Legal Information Institute (CanLII) website" to provide "versatile solutions for various use cases." By working with industry, LexUM said, it could help develop new and exciting tools:

> <u>We believe that our algorithms trained on CanLII's data have the potential to benefit your own datasets</u>. We are, therefore, extending this invitation to organizations dealing with substantial volume of legal data – legal departments, courts, tribunals,

---

[65] *A.T. v. Globe24h.com*, 2017 FC 114 (CanLII), [2017] 4 FCR 310.
[66] "The Canadian Legal Information Institute (CanLII) acquires Lexum" (28 February 2018), online: https://lexum.com/en/blog/canadian-legal-information-institute-CanLII-acquires-montreal-technology-firm-lexum.
[67] The terms of the sale are not public. However, there is information in Law Society of Ontario's and the Law Society of Alberta that can help us understand the nature of the deal. It appears that the all the law societies made an initial $2 million loan to CanLII in 2018 to fund the deal (see Law Society of Alberta, *Financial Statements (2018)* (Law Society of Albera, 2019) at 4). Each law society also committed to providing CanLII with three additional payments to finance the remainder of the term. In the Law Society of Ontario's case, this meant that the organization loaned $878,000 (43.9% of the total $2 million loan) in 2018 and made three additional payments of $280,000 in 2019, 2020, and 2021, for a total of $840,000 (see Law Society of Ontario, *Financial Statements (2018)* (Law Society of Ontario, 2019) at 21). Assuming that the Law Society of Ontario was responsible for the same proportion of balance payments as it was for the initial loan, this would mean that all the law societies contributed an additional $2 million.
[68] "Transparence.Arkansas.Gov", online: https://transparency.arkansas.gov.
[69] "The Arkansas Judiciary Adopts Decisia for the Online Publishing of its Opinions" (21 June 2017), online: https://lexum.com/en/blog/the-arkansas-judiciary-adopts-decisia-for-the-online-publishing-of-its-opinions
[70] "New Mexico Sunshine Portal", online: https://ssp3.sunshineportalnm.com

publishers, and others – to join a pilot project aiming at testing the potential of our R&D efforts for your business.[71] [emphasis added]

Over the next few years, Lexum continued to use CanLII data to develop new technologies and projects. In 2019, for example, Lexum introduced Facts2Law, a machine learning approach to predict the legal authorities that would be most relevant to a given text.[72] In 2022, announced that it was developing technology to automatically classify caselaw.[73]

Just as Lachance predicted back in 2015, others wanted to see what they could do with legal data and legal decisions. In a 2018 submission to Parliament's Standing Committee on Industry, Science and Technology, CanLII explained that inadequate access to legal data was inhibiting innovation in the legal field, writing that the institute "has direct knowledge that several Canadian startups interested in developing such solutions were discouraged due to the lack of access to data and have abandoned the legal sphere altogether."[74] In 2021, the Canadian Judicial Council published guidelines, recognizing that courts and tribunals were facing increased requests from commercial publishers for data, to help court administrators adjudicate applications for access to legal data.[75]

Shortly after the report was released, CanLII and Lexum jointly authored a blog post welcoming the report, explaining their own reluctance to provide data: "[w]e never believed we had clear permission to redistribute this content and decide who should be granted or denied bulk access to legal decisions." But if courts and tribunals officially signed off on allowing entities to access data, "as the technology supplier of decision websites for a number of the courts and tribunals, [Lexum] is ready to enable bulk access."[76] But this claims rings hollow because many courts in Canada already allow technologically sophisticated users to download their entire datasets, but neither CanLII nor Lexum have yet, as they indicated they would, help facilitate access.

---

[71] "Lexum is Inviting Industry Partners to Join Lexum Lab in Pilot Implementations of Legal Data Enhancement Technologies" (27 September 2018), online: https://lexum.com/en/blog/lexum-is-inviting-industry-partners-to-join-lexum-lab-in-pilot-implementations-of-legal-data-enhancement-technologies

[72] Ivan Mokanov, Daniel Shane, and Benjamin Cerat, "Facts2law: using deep learning to provide a legal qualification to a set of facts" (2019) *Proceedings of the seventeenth international conference on artificial intelligence and law*.

[73] Lexum, "Lexum's Approach to Automatic Classification of Case Law: From Statistics to Machine Learning" (8 April 2022), online: https://www.slaw.ca/2022/04/08/lexums-approach-to-automatic-classification-of-case-law-from-statistics-to-machine-learning/

[74] Brief submitted to House of Commons, Standing Committee on Industry, Science and Technology, *Evidence* (29 August 2019), online: https://www.ourcommons.ca/Content/Committee/421/INDU/Brief/BR10020436/br-external/CanadianLegalInformationInstitute-e.pdf

[75] Jo Sherman, *Guidelines For Canadian Courts Management of Requests for Bulk Access to Court Information by Commercial Entities* (Canadian Judicial Council, 2021).

[76] Xavier Beauchamp-Tremblay, Pierre-Paul Lemyre, Sarah Sutherland and Ivan Mokanov, "Comments on the New CJC Guidelines on Bulk Access to Court Information" (28 May 2021), online: https://www.slaw.ca/2021/05/28/comments-on-the-new-cjc-guidelines-on-bulk-access-to-court-information/

Moreover, CanLII is willing to move to prevent unauthorized use of its datasets. In November 2024, CanLII initiated an action against Caseway, a Vancouver-based startup that developed a chat-bot search tool for Canadian law. CanLII claimed that the Caseway downloaded the entire CanLII dataset in violation of the terms of service. CanLII also claimed copyright in the data, arguing that its efforts to organize the legal decisions (amongst other editorial work) gave CanLII copyright interests in the legal decisions. The case remains pending, and it is not clear what arguments CanLII will advance if the matter proceeds, but if CanLII presses its copyright claims, there is an uncomfortable echo to earlier litigation when for-profit publishers claimed that public libraries were breaching their copyright by photocopying decisions.[77]

And this is to say nothing of academic researchers. If once LexUM was a leader in free access to law committed to advancing scholarly work, it is regrettable that CanLII and Lexum do little to support the empirical, technological, and sociological research that would depend on access to large volumes of data. Once leaders, today CanLII and Lexum have narrow understandings of what free access to law could mean. At best, the fact that CanLII is not one of the strongest advocates of free access to law is a disappointment; at worst, it is a sign that the old pattern repeats: worried about protecting revenue, CanLII and Lexum see their legal databases as a strategic asset, to be managed on their own terms.

## III. THE A2AJ'S CANADIAN LEGAL DATA: AN OPEN-SOURCE ALTERNATIVE

Because CanLII is not making Canadian legal data available to researchers and developers, Access to Algorithmic Justice (A2AJ) is stepping in to provide an alternative.

The A2AJ research project – led jointly by this article's authors – represents a collaboration between York University's Osgoode Hall Law School and Toronto Metropolitan University's Lincoln Alexander School of Law. With financial backing from the Law Foundation of Ontario and infrastructure support from the Digital Research Alliance of Canada, we work to advance and promote a more equitable and accessible justice system in an era of technological transformation. Our initiatives encompass creating legal datasets, establishing benchmarks for legal AI systems, and building open-source technological solutions for the legal sector – with particular emphasis on serving Canada's marginalized and economically disadvantaged populations. We champion legislative reforms that harness legal technology to benefit the broader public rather than only those with substantial resources. Additionally, we facilitate cross-disciplinary educational experiences for students in both law and technology fields. Our overarching vision is to cultivate a Canadian legal technology landscape that prioritizes non-profit models, open-source principles, and the protection and advancement of fundamental rights.[78]

---

[77] Notice of Claim in *CanLII v 1345750 B.C. Ltd., et al.*, Supreme Court of British Columbia, VLC-S-S-247574.

## Introducing the A2AJ's Canadian Legal Data

The A2AJ's *Canadian Legal Data* (CLD) project[79] offers programmatic access to Canadian legal data through multiple open-access and open-source channels. The project builds on an earlier dataset[80] that was maintained by the Refugee Law Lab, which incubated the A2AJ, and which continues to host the infrastructure for the A2AJ. The data is intended to support empirical legal research, legal-tech prototyping, and language-model pre-training in the public interest – especially work that advances access to justice for marginalized and low-income communities.

Currently, the CLD includes the full text of over 116,000 court and tribunal decisions and over 5,000 statutes and regulations, most of which are available in both English and French.[81] Taken together, the CLD includes over 1.2 billion tokens[82] of legal text (or approximately 1 billion words).[83] Tables 1 and 2 offer further detail about current coverage.

---

[78] A2AJ, "About Us" (2025), online: <https://a2aj.ca/about>.
[79] Sean Rehaag & Simon Wallace, "A2AJ Canadian Legal Data" (2025), online: <https://github.com/a2aj-ca/canadian-legal-data>.
[80] Sean Rehaag, "Refugee Law Lab: Canadian Legal Data" (2023) online: <https://huggingface.co/datasets/refugee-law-lab/canadian-legal-data> (updated 2024).
[81] Rehaag & Wallace, supra note __.
[82] Tokenization is a method of converting words or parts or words into numbers to facilitate algorithmic calculations, including by large language models. See e.g., OpenAI, "Tiktoken" (updated 2025), online: <https://github.com/openai/tiktoken>. See also, Rico Sennrich, Barry Haddow & Alexandra Birch, "Neural Machine Translation of Rare Words with Subword Units" (2016) 2016:1 Proceedings of the 54th Annual Meeting of the Association for Computational Linguistics 1715, online: DOI: 10.18653/v1/P16-1162.
[83] Sean Rehaag, "Notebook to count the number of tokens (words) in legal datasets" (2025), online: <https://github.com/a2aj-ca/canadian-legal-data>.

Table 1: Case Law Dataset Coverage

| Dataset | Earliest | Latest | Documents | Token Count |
|---|---|---|---|---|
| Canadian Human Rights Tribunal | 2003-01-10 | 2025-07-16 | 1,050 | 21,875,457 |
| Court Martial Appeal Court | 2001-01-19 | 2025-06-17 | 147 | 2,352,315 |
| Federal Court | 2001-02-01 | 2025-08-01 | 34,256 | 409,525,860 |
| Federal Court of Appeal | 2001-02-01 | 2025-08-01 | 7,580 | 74,318,193 |
| Ontario Court of Appeal | 2007-01-02 | 2025-08-01 | 16,951 | 59,157,061 |
| Refugee Appeal Division (IRB) | 2013-02-19 | 2024-07-22 | 14,022 | 172,800,873 |
| Refugee Law Lab Reporter (RPD, IRB) | 2019-01-07 | 2023-12-29 | 898 | 2,459,030 |
| Refugee Protection Division (IRB) | 2002-07-16 | 2020-12-14 | 6,729 | 79,241,147 |
| Supreme Court of Canada | 1877-01-15 | 2025-07-31 | 10,845 | 179,880,823 |
| Social Security Tribunal | 2013-03-08 | 2025-12-16 | 16,338 | 119,870,547 |
| Tax Court of Canada | 2003-01-21 | 2025-07-30 | 7,918 | 109,061,699 |
| TOTAL | | | 116,734 | 1,230,543,005 |

Table 2: Laws Dataset Coverage

| Dataset | Earliest | Latest | Documents | Token Count |
|---|---|---|---|---|
| Federal Statutes | 1870-05-12 | 2025-06-26 | 954 | 28,692,454 |
| Federal Regulations | 1945-12-21 | 2025-07-16 | 4,803 | 26,884,785 |
| TOTAL | | | 5,757 | 55,577,239 |

Our *Canadian Case Law*[84] and *Canadian Laws*[85] datasets both have similar structures, with the one exception being that for the Canadian Laws dataset we provide access both to the full text of regulations and legislation, as well as to structured text that allows for access to specific sections. Table 3 describes the available fields for the datasets.

Table 3: Dataset Fields

| Field | Description |
|---|---|
| dataset | Abbreviation identifying the court/tribunal/jursidiction |
| citation_en / citation_fr | Citation in English / French (neutral citation where available) |
| citation2_en / citation2_fr | Secondary citation(s) where available |
| name_en / name_fr | Name of the document (style of cause where case) |
| document_date_en / document_date_fr | Date of the document (decision date where case) |
| url_en / url_fr | Source URL (or where URL not available, other identifier) |
| scraped_timestamp_en / scraped_timestamp_fr | Timestamp when the document was obtained |
| unofficial_text_en / unofficial_text_fr | Full unofficial text of the document |
| unofficial_sections_en / unofficial__sections_fr | Parsed sections of legislation / refgulations (laws dataset only) |
| upstream_license | License terms from the data source |

# Data Collection & Licensing

As of the time of writing, the A2AJ obtains data for the CLD through three primary means. First, we use automated processes to scrape the websites of courts and tribunals that make their decisions available online[86] – with further processes in place to update the data daily via RSS feeds[87] and webpages that list recent decisions where RSS feeds are not available.[88] Second, we obtain some decisions directly from relevant administrative tribunals through the same email distribution processes available to legal publishers.[89] Third, we obtain data about Federal legislation and regulations from a comprehensive code repository maintained by the Department of Justice, and we check for updates on a weekly basis.[90]

---

[84] Sean Rehaag & Simon Wallace, "A2AJ Canadian Case Law" (2025), online: <https://huggingface.co/datasets/a2aj/canadian-case-law> [Rehaag & Wallace, "HF Case Law"].
[85] Sean Rehaag & Simon Wallace, "A2AJ Canadian Laws" (2025), online: <https://huggingface.co/datasets/a2aj/canadian-laws> [Rehaag & Wallace, "HF Laws"].

We are careful to comply with terms of service of the websites that we scrape. Many of those websites include limitations in terms of service, either in general or specifically with regard to decisions on their websites.[91] Our dataset includes details about those terms of service for each document we reproduce, and users are required to comply with these upstream licenses.

Because of our commitment to complying with terms of service of our data sources, we currently do not scrape CanLII, which significantly restricts the scope of our data. Because many courts and tribunals do not publish decisions on their websites and instead publish only on CanLII (and via commercial legal publishers) our dataset is not yet comprehensive.[92] We are hoping to expand our access to data by establishing processes to obtain decisions directly from additional courts and tribunals – beginning with tribunals that operate in areas of law that disproportionately impact low income communities, such as the Landlord Tenant Board and the Human Rights Tribunal of Ontario. We also hope either to persuade CanLII to adopt less restrictive terms of service or to convince more courts and tribunals to host decisions on their own websites if CanLII refuses to do so.

---

[86] See e.g. Supreme Court of Canada, "Decisions and Resources" (last updated 2025), online: <https://decisions.scc-csc.ca/scc-csc/en/nav.do>.

[87] See e.g. Supreme Court of Canada, "Supreme Court Judgments" (last updated 2025), online: <https://decisions.scc-csc.ca/scc-csc/scc-csc/en/rss.do>.

[88] See e.g., Federal Court of Appeal, "New Decisions" (last updated 2025), online: <https://decisions.fca-caf.gc.ca/fca-caf/en/ann.do>.

[89] For example, we work with our partner organization, the Refugee Law Lab (RLL), to obtain Immigration and Refugee Board decisions directly from the Immigration and Refugee Board via the same email distribution channels used by the Immigration and Refugee Board to send decisions for publication to other legal publishers, such as CanLII. We also reproduce cases published in the Refugee Law Lab Reporter, which the RLL obtains via access to information requests. Refugee Law Lab, "Refugee Law Lab Reporter" (last updated 2025), online: <https://refugeelab.ca/rllr/>.

[90] Department of Justice, "laws-lois-xml" (last updated 2025), online: <https://github.com/justicecanada/laws-lois-xml>.

[91] See e.g., Supreme Court of Canada, "Terms and Conditions" (last updated 2025), online: <https://scc-csc.ca/resources-ressources/terms-avis/#copyright-and-permission-to-reproduce>.

[92] See e.g., Alberta Court of Appeal, "Judgments" (last updated 2025), online: <https://www.albertacourts.ca/ca/publications/judgments>; Ontario Superior Court of Justice, "Decisions of the Court" (last updated 2025), online: <https://www.ontariocourts.ca/scj/about-the-court-2/decisions-of-the-court/>; Tribunals Ontario, "Law, Rules and Decisions" (last updated 2025), online: <https://tribunalsontario.ca/ltb/law-rules-and-decisions/#decisions>.

Also, because we pass along the upstream licenses through which we obtain each document, while our overall code and dataset is open source, many individual documents have limitations on use, including in some cases non-commercial use restrictions. We are actively working to get permission from courts and tribunals to remove these limitations to make our data less restrictive. We plan to begin with Federal courts and tribunals, which are subject to the regulation noted earlier that allows for the reproduction of court and tribunal decisions, "provided due diligence is exercised in ensuring the accuracy of the materials reproduced and the reproduction is not represented as an official version."[93]

## Accessing the A2AJ's Canadian Legal Data

The A2AJ's Canadian Legal Data is designed to be as accessible as possible, recognizing that different users have different technical capabilities and use cases. We provide four methods for accessing the data, each optimized for particular needs and technical contexts. All access methods are free of charge and require no registration, embodying our commitment to removing barriers to legal data access.

### API Access for Targeted Searches

For users who need to search for specific cases or laws, or who want to integrate legal data into applications, we provide a RESTful Application Programming Interface (API).[94] This interface allows users to programmatically search by citation, name, full text, date ranges, and jurisdiction. The API returns structured JSON data[95] that can be easily integrated into websites, legal research tools, or analytical workflows. Full documentation and interactive examples are available, including in notebook tutorials demonstrating common use patterns.[96]

### Hugging Face Datasets for Machine Learning Research

Recognizing the growing importance of natural language processing in legal technology, we host our bulk datasets on Hugging Face, the leading platform for machine learning datasets.[97] The *Canadian Case Law* dataset[98] and *Canadian Laws* dataset[99] can be accessed directly through the Hugging Face datasets library, making it trivial for researchers to load the data for training language models, conducting computational legal studies, or developing AI-powered legal tools. This method is particularly valuable for academic researchers and non-profit organizations developing open-source legal AI systems that can benefit underserved communities. Example code for how to load the datasets are available in notebooks.[100]

---

[93] *Reproduction of Federal Law Order*, SI/97-5.
[94] Access to Algorithmic Justice, "Canadian Legal Data API" (last updated 2025), online: <https://api.a2aj.ca/docs>. See also, Github, "What is an API?" (7 July 2025), online: <https://github.com/resources/articles/software-development/what-is-an-api>.
[95] Anca-Raluca Breje, et al, "Comparative Study of Data Sending Methods for XML and JSON Models" (Comparative Study of Data Sending Methods for XML and JSON Models (2018) 9:12 International Journal of Advanced Computer Science and Applications 198 (explaining what JSON is and how it works).
[96] Access to Algorithmic Justice, "A2AJ Canadian Legal Data API" (last updated 2025), online: <https://github.com/a2aj-ca/canadian-legal-data/blob/main/access-via-api.ipynb>.
[97] Hugging Face, "Hugging Face – The AI community building the future" (last updated 2025), online: <https://huggingface.co/>.
[98] Rehaag & Wallace, "HF Case Law", supra note ___.
[99] Rehaag & Wallace, "HF Laws", supra note ___.

### Direct Parquet Downloads for Bulk Analysis

For users who need the entire dataset but prefer minimal technical dependencies, we provide direct downloads in the Parquet format – an efficient columnar storage format that can be read by most data analysis tools including R and Python.[101] This approach serves researchers conducting large-scale empirical studies of Canadian law, such as analyzing outcome patterns, tracking legal precedent evolution, or studying access to justice metrics across different jurisdictions and time periods. As with other ways to access the data, we provide a notebook to demonstrate how to access the data.[102]

### Model Context Protocol for AI Integration

As generative AI becomes increasingly important for legal assistance, we've implemented the Model Context Protocol (MCP)[103] – which is a recently established standardized communication protocol that facilitates use of data and other resources by AI agents.[104] Through our MCP, AI assistants and chatbots can easily query our legal database in real-time, enabling them to provide accurate, up-to-date Canadian legal information. Community legal clinics could use this to power chatbots that help clients understand their rights, while pro bono lawyers could use AI assistants with direct access to relevant case law. We have put together a notebook that shows how to integrate our MCP into common AI platforms – including through programmatic means for users with coding skills or via web interfaces for users without those skills who simply want to connect web-based AI tools that they are already using with our data.[105]

Each access method includes comprehensive documentation and example code. We've deliberately chosen widely supported, open standards to ensure the data remains accessible regardless of users' technical resources or expertise. This multi-channel approach reflects our belief that legal data should be available to everyone – from individual self-represented litigants using simple web searches, to legal aid organizations building sophisticated analytical tools, to researchers advancing our understanding of how law operates in practice.

---

[100] Access to Algorithmic Justice, "A2AJ Canadian Legal Data Hugging Face Datasets" (last updated 2025), online: <https://github.com/a2aj-ca/canadian-legal-data/blob/main/access-via-hugging-face.ipynb>.
[101] Apache, "Apache Parquet" (last updated 2025), online: <https://parquet.apache.org/>.
[102] Access to Algorithmic Justice, "A2AJ Canadian Legal Data Downloaded Via Parquet Files" (last updated 2025), online: <https://github.com/a2aj-ca/canadian-legal-data/blob/main/access-via-hugging-face.ipynb>.
[103] Access to Algorithmic Justice, "MCP" (last updated 2025), online: <https://api.a2aj.ca/mcp>.
[104] Model Context Protocol, "Get started with the Model Context Protocol (MCP)" (last updated 2025), online <https://modelcontextprotocol.io/docs/getting-started/intro>. See also, Anthropic, "Introducing the Model Context Protocol" (4 November 2024), online: < https://www.anthropic.com/news/model-context-protocol>.
[105] Access to Algorithmic Justice, "A2AJ Canadian Legal Data via MCP" (last updated 2025), online: <https://github.com/a2aj-ca/canadian-legal-data/blob/main/access-via-mcp.ipynb>.

## Example Use Cases for A2AJ's Canadian Legal Data

To demonstrate the potential of the CLD, we present two examples that showcase different use cases: supporting institutional efficiency and accessibility initiatives within courts, and developing automated legal intelligence tools.

### Example 1: Supporting Efficiency and Accessibility Initiatives

Courts and tribunals across Canada are increasingly focused on improving both the accessibility of their decisions to the public and the efficiency of their processes. In this context, the CLD can helpfully enable courts and tribunals to conduct data-driven assessments of their own practices and measure the impact of institutional reforms.

Consider a scenario where the Tax Court of Canada launches a one-year initiative to improve the accessibility of their written decisions by holding a series of judicial education seminars on readability. If the Court wanted to measure the success of that initiative, they could easily use the CLD to compare automated readability scores using the Flesch Reading Ease metric for decisions written in the year before and the year after the training. A notebook available on the A2AJ Canadian Legal Data GitHub repository shows how to use the CLD to generate a chart with the median word count of Tax Court of Canada decisions in just a few lines of code:[106]

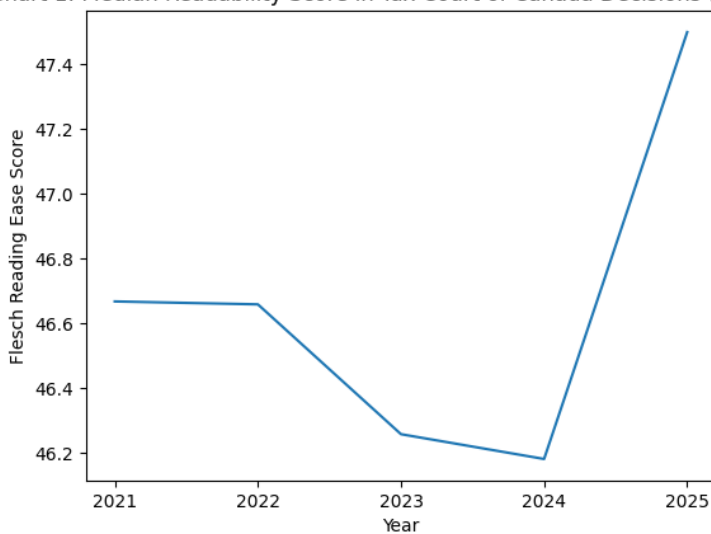

---

[106] Sean Rehaag, "Notebook to examine word count and readability in court decisions" (2025), online: <https://github.com/a2aj-ca/canadian-legal-data/example-projects/readability.ipynb>.

Or imagine a scenario where the Federal Court seeks to identify best practices in efficient decision writing in immigration judicial reviews, which represent the Court's highest volume area. Using the CLD, they could quickly analyze all immigration decisions from a given period to identify judges who consistently produce concise decisions and have those judges provide training to their colleagues about judicial decision-writing. The same notebook on the GitHub repository provides code to calculate the median word counts for all Federal Court justices in immigration judicial reviews from 2021 to 2025. That code produces this table, which reveals significant variance in median word count across justices:[107]

Table 4: Median Decision Word Count by Federal Court Justice (Immigration Files, 2021-2025) (5 Highest and Lowest)

| Justice | Median Word Count | Decisions |
| --- | --- | --- |
| HENEGHAN | 806 | 190 |
| BATTISTA | 1,190 | 70 |
| GRAMMOND | 1,213 | 102 |
| SADREHASHEMI | 1,563 | 201 |
| MCDONALD | 1,911 | 104 |
| LITTLE | 3,954 | 143 |
| BROWN | 4,135 | 139 |
| GASCON | 4,152 | 88 |
| ROY | 4,322 | 77 |
| STRICKLAND | 4,719 | 106 |

Of course, our point here is not about the metrics chosen for these examples. For instance, the Flesch Reading Ease may not necessarily be the best measure of readability in legal texts. Some scholars criticize this measure and instead propose alternative custom metrics.[108] Similarly, one might take issue with the presumed connection between shorter decisions and efficiency. For example, one might argue that there are important tradeoffs when judges do not take the time to fully explain their reasoning, or that patterns in case assignment or other practices drive variance in decision word counts rather than efficiency.

Our point is simply that, as these two examples show, the CLD facilitates easy application of empirical metrics to legal text in a way that could help courts, tribunals, and other institutions gain further insight into their operations – and ultimately take evidence-informed measures to attempt to improve access to justice.

### Example 2: Automated decision-summarizer and podcast generator

Advocates need new ways to keep track of first-instance jurisprudence. We live in times of jurisprudential scale. Consider the case of an immigration and refugee lawyer. Every week, the Federal Court of Canada releases judicial review decisions that help advocates and government officials understand how immigration and refugee decisions ought to be made. As Chart 2 shows, the quantity of text concerning immigration and refugee decisions released by the Court is large and growing. In the median week in 2000, the Court released under 30,000 words a week; in the median week in 2024, the Court released almost 50,000 words.

---

[107] Ibid.
[108] See e.g., Mike Madden, "How Understandable are Adjudicative Decisions? Introducing and Applying Law's Own Readability Formula" (forthcoming) 30 Legal Writing, draft available online: <https://ssrn.com/abstract=5110351>.

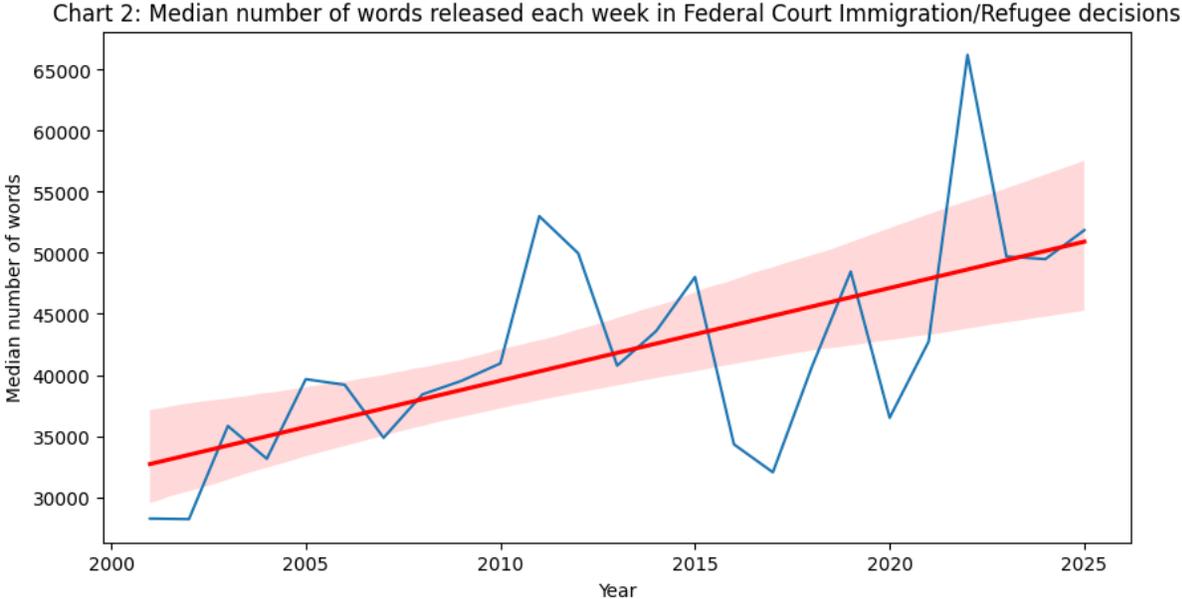

For the overworked legal aid lawyer or sole practitioner, this is just too much to read and too much to stay on top of. And, of course, it is a shame. Effective advocacy is not just knowing Supreme Court of Canada precedents, it is also knowing about intermediate level debates, the sorts of problems that regularly crop up, the ways in which those debates are considered, and the emergence of novel issues.

The A2AJ project now allows developers to build novel current awareness tools. For example, in a few lines of code, we can make a tool that reads the weekly immigration and refugee law output of the Federal Court, identifies decisions where the Court found an error, and produce a digestible memorandum that organizes cases by the application type and identifies the error. As we can see below, without much effort, it is now possible to produce a regularly updating memo.

```
Memorandum: Federal Court immigration/refugee judicial reviews decided last week

Overview
- Total FC immigration/refugee decisions: 13
- Judicial reviews allowed: 3
- Total words released: 24,264

Key themes: The successful reviews turned on unreasonable assessments of the
evidence and inadequate reasons. In each case, the decision-maker either
misapprehended key facts, failed to grapple with central evidence, or relied on
a flawed characterization that drove the outcome. One case also rejected
post-hoc rationalizations offered in litigation.

Case summaries

1) Ahmed v. Canada (Citizenship and Immigration), 2025 FC 1449 (Grammond J.) —
Economic (work permit refusal; consequential visitor/permit refusals)
- Facts: A Pakistani applicant with a bachelor's degree in accounting and
finance sought a work permit as an accounting technician/bookkeeper. The officer
refused, citing lack of work experience and English ability and concluding he
would not leave Canada. At the hearing, the Minister conceded that because of
the applicant's university degree, work experience was not required for the NOC
occupation.
- Errors: The officer unreasonably focused on supposed gaps in work experience
(and deficiencies in reference letters) even though experience was not required
```

Outputs do not need to be in memo form either. As we show in a notebook[109] in our GitHub repository, we can transform this memorandum into a podcast script and use modern artificial intelligence tools to transform that text into speech.[110]

The most remarkable feature of this little example, and of the A2AJ, is that the entire tool – that identifies recent decisions, that finds the ones where the error was identified, that produces a memo, and that then makes a podcast – took only two hours to build. While a close review shows real room for improvement, the availability of A2AJ data now makes iteration and development considerably easier.

---

[109] Simon Wallace, "Notebook to make a weekly summary of immigration and refugee law jurisprudence" (2025), online: https://github.com/a2aj-ca/canadian-legal-data/example-projects/fc_summarizer.ipynb.
[110] You may listen to the sample podcast here: https://github.com/a2aj-ca/canadian-legal-data/blob/main/example-projects/podcast.mp3

# CONCLUSION: NEXT STEPS FOR THE A2AJ'S CANADIAN LEGAL DATA

Canada was once a leader in free access to law, but it is not any longer. As the power of new computational research techniques is starting to come into view, CanLII ought to be leading efforts to democratize access to legal data. Its failure to do so is already consequential. Huge multi-billion-dollar corporations are investing enormous sums in tools and datasets that only the well-resourced will have access to. If it is not already the reality today, it will be tomorrow: the wealthy are about to get better legal services because they can pay for computational tools built on proprietary legal data. Meanwhile, CanLII – which Canadian lawyers have financed for years – continues to deny public and scholarly access to one of the largest collections of Canadian legal data.

This is why we established the Access to Algorithmic Justice Project. Despite many efforts, Canada has struggled for a century and a half to increase the public's access to legal data and legal information. Too often, inspiring efforts failed when private interests saw that a buck was to be made. As a growing digital divide comes into view, we know how urgent this project is. By open-sourcing legal data, undertaking research, providing training and infrastructure, and building new legal research tools, our hope for the A2AJ is to spur and support a vibrant computational access to justice ecosystem in Canada.

But this is our second-best outcome because we should not need to duplicate CanLII's work. Our preference would instead be to focus our efforts on research, on development, and on training. We hope that, once this project becomes better known, governments and the legal profession will agree that it makes no sense for us to do what CanLII ought to do. The best projects share a common goal: they hope to make themselves redundant. Fortunately, we think believe that redundancy *could* be achieved soon if action is taken soon.

The law societies and the Federation of Law Societies must apply immediate pressure on CanLII. CanLII is rightly celebrated as a major achievement, but for it to continue to succeed in an era of computational law it needs to re-assume its leadership position. If CanLII prefers to maintain the *status quo*, it is more likely that the entire computational legal space will be ceded to well-resourced actors.

Courts and tribunals must recognize that access to data is a key component of access to justice. Many Canadian courts and tribunals have already taken major and significant efforts to ensure that their courts are open, not just physically, but also digitally. These efforts are commendable and should inspire projects across the justice system. The law does not belong to anyone, it belongs to the public. Courts and tribunals must ensure that legal data is not privatized for profit, but made available as a public good. Too often courts and tribunals are cited as the reasons why data cannot be made freely available. Going forward the judiciary should assume a clear and unambiguous posture: that it supports free access to law.

Twenty-five years ago, CanLII was established with a mandate to increase access to law and ensure that legal information would not just benefit well-resourced actors. We made the A2AJ to protect this legacy and to head off the coming digital divide. With luck, we are just a course correction away from finding ourselves in good company.